# Comment on "Ballistic Majorana Nanowire Devices" by Gül et al. Nature Nanotechnology 2018

Sergey Frolov, University of Pittsburgh
Vincent Mourik, FZ Jülich

July 28, 2024

## Summary

This work re-analyzes Gül et al. Nature Nanotechnology 2018 "Ballistic Majorana nanowire devices" using fuller data from the original experiments released in 2023 on Zenodo. The authors have prepared a correction to their article that appeared in Nature Nanotechnology in 2024. However, the correction does not address the concerns we identify here. We demonstrate that the fuller data contain extensive evidence for quantum dots and disorder that are completely inconsistent with the authors' conclusion that they have achieved ballistic devices containing zero bias peaks of likely Majorana origin. We show how data selection, data cropping and undisclosed data processing played a role in composing the figures of the final published paper.

## Why the use of "ballistic" is central

The central claim of the paper "Ballistic Majorana Nanowire Devices" by Gül et al, Nature Nanotechnology 2018[1] is contained in the title of the paper. The authors use the term "ballistic" to assert that their samples are of sufficient quality to plausibly observe Majorana modes and to lay the groundwork for these claims. The exact criteria for what the authors mean by their term "ballistic" are given by the authors in the text of this paper, and in the text of their 2017 Nature Communications paper titled "Ballistic superconductivity in semiconductor nanowires" [2]. In both papers, and in the original manuscript that both came from[3], the term "ballistic" communicates that the nanowires exhibit quantized conductance due to one-dimensional mode-resolved transport, and that undesirable quantum dots are absent from the nanowires.

**Ballistic Majorana nanowire devices**

Önder Gül[1,5*], Hao Zhang[1*], Jouri D. S. Bommer[1], Michiel W. A. de Moor[1], Sébastien R. Plissard[2,6], Erik P. A. M. Bakkers[1,2], Attila Geresdi[1], Kenji W Takashi Taniguchi[3] and Leo P. Kouwenhoven[1,4*]

Majorana modes are zero-energy excitations of a topological superconductor that exhibit non-Abelian statistics[1–3]. Following proposals for their detection in a semiconductor nanowire coupled to an s-wave superconductor[4,5], several tunnelling experiments reported characteristic Majorana signatures[6–11]. Reducing disorder has been a prime challenge for these experiments because disorder can mimic the zero-energy signatures of Majoranas[12–16], and renders the topological properties inaccessible[17–20]. Here, we show characteristic Majorana signatures in InSb nanowire devices exhibiting clear ballistic transport properties. Application of a magnetic field and spatial control of carrier density using local gates generates a zero bias peak that is rigid over a large region in the parameter space of chemical potential, Zeeman energy and tunnel barrier potential. The reduction of disorder allows us to resolve separate regions in the parameter space with and without a zero bias peak, indicating topologically distinct phases. These observations are consistent with the Majorana theory in a ballistic system[21] and exclude the known alternative explanations that invoke disorder[12–16] or a nonuniform chemical potential[22,23].

superconductor (purple) and no local bottom gate electrodes are s boron nitride flake and are opera tial control of the carrier density i our devices following our recently which results in a high-quality In hard superconducting gap, and b tized nanowire (see refs [19,20]). All r dilution refrigerator with an elect data is taken by applying a bias vol lead and the superconductor indic monitoring the current flow. The c

Figure 1b shows the differe varying V, and stepping the vol Importantly, we find no signs o any other localization effects. Ver indicated with coloured bars are tom) is from the tunnelling regime negative voltage on the barrier gat nanowire section, and creates a tu lead and the superconductor. In

The claim of ballistic devices is central to the paper

"Here, we show characteristic Majorana signatures in nanowire devices that **exhibit ballistic transport**, ruling out all known disorder- or nonuniformity-based explanations."

"We have realized our devices following our recently developed nanofabrication recipe which results in a high-quality InSb–NbTiN interface, an induced hard superconducting gap, and **ballistic transport** in the proximitized nanowire (see refs [19,20])."

They refer to Zhang Nature Comm 2017, which is another paper that raises concerns of similar type

We analyze the claims of the paper in the context of the field as it was back in 2015-2018 when the authors performed and published their experiments. Although our analysis remains valid today, we do not rely on any subsequent scientific insights that were unknown to the authors in 2015. Already in 2015, Majorana experts including these authors knew that a key signature for the formation of Majorana zero modes, the zero bias conductance peak, could easily have another origin[4]. It can appear due to disorder-induced unintended quantum dots, in which trivial states, often referred to as Andreev bound states, exist beyond the experimentalists' control. When magnetic fields are applied, these trivial states congregate near zero bias forming zero bias peaks in tunneling conductance, that superficially resemble Majorana modes, without the Majorana physics present.

To publish a paper interpreting zero bias peaks as Majorana zero modes the authors argued, in the first part of the paper, that their nanowires were free of quantum dots ("ballistic") thereby countering the most common explanation for zero bias peaks – that they are due to trivial states. Having set the stage this way, they proceeded in the second part of the paper to show various supposedly robust zero bias peaks that they claim are due to Majorana. In other words, it is both the absence of quantum dots and the appearance of supposedly robust zero bias peaks that renders the authors' conclusions plausible.

The authors of the original paper have published a correction in Nature Nanotechnology in 2024[5]. This correction does not take care of the issues we identify here, and the authors appear to stand by their claims. For example, corresponding author and lead PI of the study Leo Kouwenhoven has renewed his praise of the data from this Nature Nanotechnology paper. He features these data in Figure 2 of his perspective posted on arxiv in June 2024[6], where he states:

*"Figure 2 taken from Ref. 32 shows one of our best results on InSb-NbTiN. The ZBP extends over more than a meV in Zeeman energy, which is more than 20 times the linewidth of the ZBP. Observations like this, data taken in 2016, made us conclude that these ZBPs did not originate from accidental crossings of Andreev bound states [33]. Also, such robust ZBPs did not seem to be expected for disordered wires and phenomena like anti-weak localization [34]. We felt in 2018 that 'all known alternative explanations, other than a Majorana explanation' were excluded by these results [32]."* Leo Kouwenhoven, "Perspective on Majorana bound-states in hybrid superconductor-semiconductor nanowires" [6] (June 24, 2024)

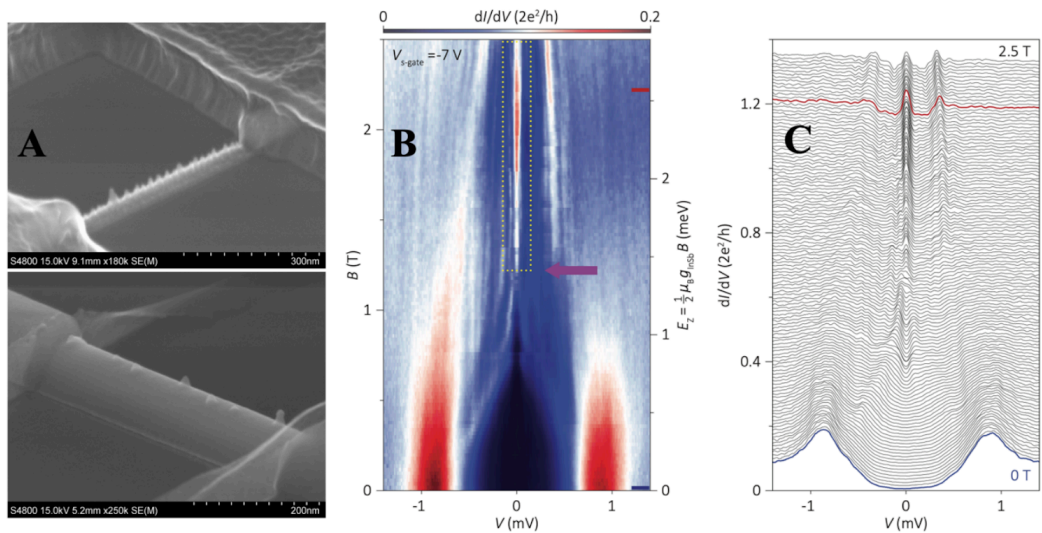

Figure 2. Top Left: Too much Argon milling of InSb nanowires can leave behind In droplets. Bottom Left: Gentle milling leaves a clean surface but still with an occasional droplet. Right panels show exemplary ZBP data on InSb-NbTiN in color scale and as line-cuts. From Ref. 32.

*Reproduced from [6] by Leo Kouwenhoven (2024)*

Re-analysis Based on 2023 Data Release

The present analysis goes beyond our 2021 analysis, shared at the time with the authors and the journal Nature Nanotechnology. We incorporate that analysis, which focused on Figures 1 and S5, into this comment and go beyond. Already in 2021 using data available at the time we demonstrated that the studied devices fail the authors' own definition of "ballistic" when the data they collected at zero magnetic field are considered. We showed that their claims of "conductance quantization" or "absence of quantum dots" were unreliable and the relevant figures were obtained through data cropping and arbitrary recalibration of offsets. In our view, already on this basis, the claim of Majorana is no longer supported because quantum dots can explain the zero-bias peaks.

The parallel publication Zhang et al, "Ballistic Superconductivity in semiconductor nanowires' Nature Communications 2017[2] contains similar issues when it comes to fuller data revealing evidence of quantum dots, in addition to data cropping, recalibration of offsets and non-standard data processing. We analyzed most data from that paper in 2022[7]. One of us, Mourik, is a co-author of that paper, he has requested to be removed from it.

Both of us, Frolov and Mourik, are former members of the Majorana project in Delft. We have first-hand knowledge about the sample preparation procedures, measurement equipment, measurement protocols. We can understand and plot the data. We were however previously unable to comprehensively assess the zero bias peak evidence in the two "ballistic" publications. The 2021 data release contained only limited finite magnetic field data. In late 2023, Delft University of Technology shared more data via the same public Zenodo repository[8]. Version 1 of the repository is the 2021 release, and the current version 3 is the 2023 release.

## Summary of the problems

Trivial states in quantum dots were known in 2015 to be capable of forming long-lived robust zero-bias peaks not of Majorana origin. The 2023 dataset contains several magnetic field series of super-gate scans that reveal ubiquitous disorder that generates unintended quantum dots. Among other things, what those data reveal is that these devices are not better than some of the earlier generations of devices made in the group.

The data released in 2023, not shown in the Nature Nanotechnology 2018 paper, contain finite magnetic field tunneling spectroscopy data on four devices, including the flagship sample, data from which were used to generate Figures 1, 2, and 3 of the paper. Reviewed together with the data represented in the original publication and contained in the 2021 data release, we find problems falling in two classes: 1) data cropping and omissions, which serve to eliminate relevant data that contradict or complicate the simplistic claims made in the paper and 2) questionable visualization techniques, including panel stretching and colorscale alteration to emphasize the features that support the hypothesis of Majorana and conceal features that go against it.

Regarding cropping, in this comment we focus on problems caused by the authors cropping out ranges of gate voltages, magnetic fields, and/or bias voltages, reducing the full data set as recorded from the measurement equipment, thereby leaving out important features that disprove the claim of ballisticity or the claim of Majorana modes. Regarding data omission, we point out that the authors have selected a few favorable looking datasets, not showing many others that not simply undermine but disprove their claims - data from the same physical samples.

Regarding non-standard visualization techniques, which resulted in striking figures used to illustrate various claims, we focus on two classes of problems. First, we find adjustments to the default colorscale of the plotting software to make less visible features contradicting either the notion of ballisticity or the notion of Majorana modes. By itself, adjustment of colorscale is routine in the presentation of data, and is acceptable when the goal is to reveal features rather than conceal them. Colorscale adjustment is not acceptable, in our view, when the goal is to hide features that undermine the stated conclusions. Second, we also show evidence of aspect ratio manipulation through stretching of panels, which serves to exaggerate how long-lived, i.e. robust to parameter changes, the zero bias peaks were from the visual point of view. This is assisted by oversampling of the axis, i.e. stepping the gate voltage or magnetic field at unnecessarily small intervals to exaggerate how smooth the curves are. Again, such oversampling could be harmless on its own, but not when the effect is to obscure unwanted effects. These factors worked together to create the appearance in the authors' data of an extraordinarily smooth background ("clean", "ballistic") upon which their supposed zero bias peaks emerged, and concealed the disordered and intermittent, quantum dot-like nature of the background features.

This cropping and omission of data, assisted by non-standard visualization techniques, result in inaccurate representation of research outcomes.

## Analysis of the Main Device 1

The authors begin their paper by presenting data from a single device, which we shall refer to as "Main Device 1."  Main text Figures 1, 2, 3 and Supplementary Information figures 1, 2 were created using the data that originated from this device. The authors' internal naming scheme for this device labels it 'Dev3' from a cooldown in October 2016.

### Figure 1

The authors use Main Device 1 to promote the Majorana origin of zero-bias peaks shown in Figures 2 and 3. They state:

*"Importantly, we find no signs of formation of quantum dots or any other localization effects."*

*"From the absence of quantum dots, the observed induced gap with a strongly reduced subgap density of states, high interface transparency, and quantized conductance, we conclude a very low disorder strength for our device[...]."*

a

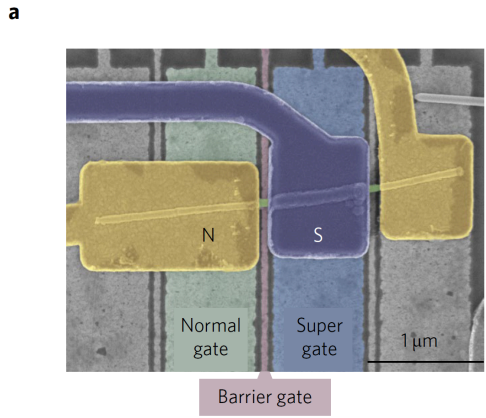

*Figure 1a reproduced from the Nature Nanotechnology 2018 paper to illustrate the device geometry. Bias voltage is applied between the N and S ("Normal" and "Superconductor") electrodes. The nanowire is clamped underneath the electrodes, and gates are underneath the nanowire. Magnetic field plots that we show feature field oriented or rotated in the plane of this image. Reproduced from [1]*

We already understood in 2021 that this device is not a ballistic device, but a device that contains evidence of quantum dots and non-quantized conductance "plateaus". We identified data manipulation in Figure 1b which contributed to the appearance of a cleaner less disordered regime at zero applied magnetic field. Data for Figure 1b were cropped: the full version of the figure demonstrates an additional resonance feature which appears similar to a quantum dot, both broadened resonances are marked by orange arrows below.

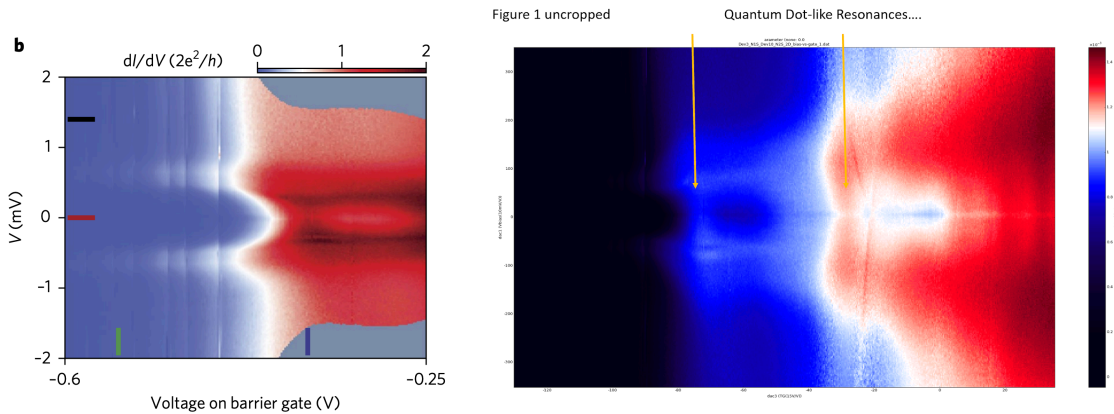

Left: Figure 1b as it appears in the paper, reproduced from[1]. Right: Data underlying Figure 1b presented without cropping and colorscale adjustment. It shows vertical features consistent with open quantum dots.

In addition to the issues above in Figure 1b, Figure 1d also underwent undisclosed processing. It exhibits a feature in conductance as a function of barrier gate voltage which the authors claim is a "*quantized conductance plateau at $2e^2/h$, a clear signature of a ballistic device.*" The conductance value at this "plateau" was subjected to an undisclosed processing step, a subtraction of series resistance of 3 kOhms. Without this, the value of conductance on the plateau would appear lower than $2e^2/h$ by a significant fraction. Further, an error in conductance calibration has been made that resulted in conductance being off by approximately 10%. The value of contact resistance, 3 kOhm, is not motivated in the paper. In fact, in all other figures, the authors subtracted a different, smaller contact resistance of 500 Ohms, also without justification.

Figure 2

Figure 2 represents the robustness of the zero bias peak to variations in magnetic field and gate voltage. This is achieved through: (1) addition of a Zeeman energy axis on the right edge of the magnetic field dependent data in panel a) to claim robustness to extreme variations in magnetic field, and (2) assertions in the caption that the data exhibits a zero bias peak robust to changing background conductance by a factor of 5 for panel b) and one order of magnitude for panel c), suggesting the zero bias peak is not affected by gates away from the superconducting device section. In contrast, the supergate dependence shown in panel d) and its corresponding description claim the supergate tunes the presence or absence of the zero bias peak. All these suggestions support the Majorana interpretation of these data.

One of our most universal findings is that the picture depicted in Figure 2 (which we reproduce in full further down this document) is non representative of the fuller data. In the 2023 data release, we found a field dependent series omitted from the paper, which includes fuller data for settings used in Figure 2. As an illustration, consider the scan below of the "Super gate," the gate underneath the superconductor S, for "Device 1," obtained at a field of 1.1T, as opposed to the 0.7T selected for Figure 2. The "Super gate" controls the part of the device where Majorana modes would, if present, be located, underneath the superconductor electrode. In fact, disorder in that part of the device is the most relevant to Majorana claims, as opposed to any disorder in the part outside the superconductor that is present in data from Figure 1, or Figures 2b and 2c.

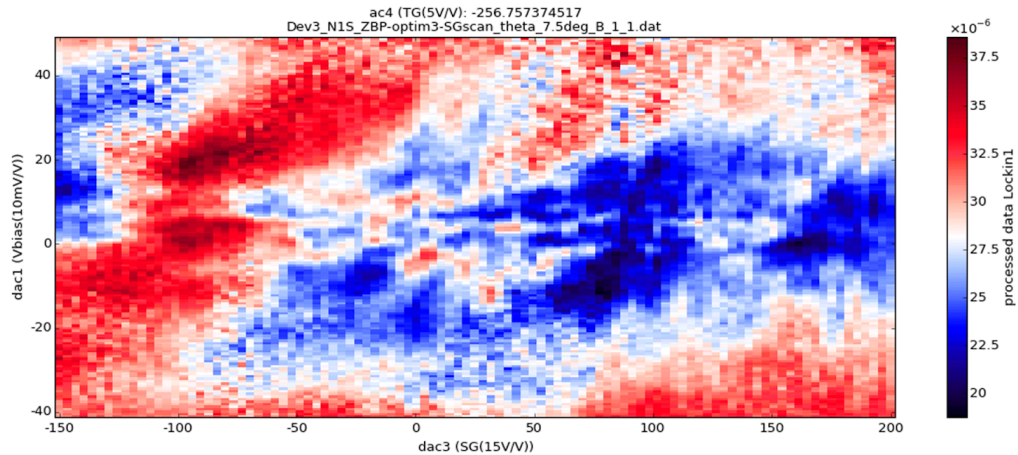

*Additional dataset shared in 2023, plotted using the default settings of the author's plotting software. Identical settings as used in Figure 2 of publication, but at different magnetic field value of 1.1T. Note that the axis labels and scales as contained in the raw files, require multipliers to be compared to paper Figures. For example, for this plot the y axis values need to be multiplied by 10 to get microvolts of bias. The x-axis values need to be multiplied by 15 to get millivolts on the gate. There is also an offset in Vbias of tens of microvolts positive, where the true zero bias is positioned. This y-offset is not a concern.*

The figure reveals a highly disordered signal, with states criss-crossing the image, forming blobs and loops. Features like this tend to arise when not just a single quantum dot, but multiple quantum dots are simultaneously present near the tunnel junction, and current flows through several of them revealing their trivial bound states. Looking at this plot, it is difficult to imagine that this is a "ballistic" device. Yet all we did to reveal this disorder is to examine a scan that is zoomed out from a narrower strip along the middle of the image that contains zero bias peaks, expanding the gate voltage range relative to that the authors selected for their Figure 2 (gate range of Fig. 2d is 40% of the gate range in the panel we selected from the fuller data).

Accordingly, we can compare this fuller data to what we find in the paper itself, as published in Nature Nanotechnology 2018. Even a visual comparison can be used to appreciate the stark difference. Figure 2 contains a clean set of images that does not show any evidence of the quantum dots and trivial states that are in fact dominating this device. Yet, our example and Figure 2 are from the same device and the same regime of parameter settings, showing data obtained almost concurrently.

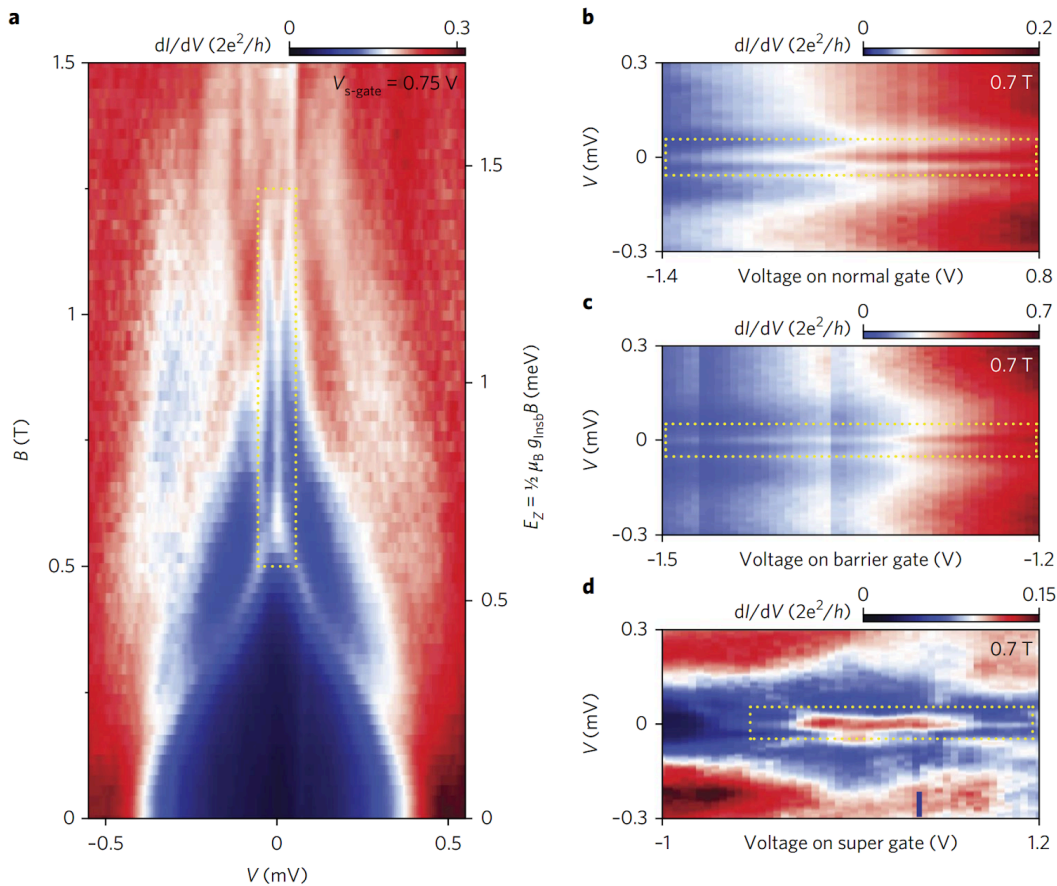

*Figure 2 reproduced from Gül et al, Nature Nanotechnology 2018 [1]*

To create Figure 2d, which illustrates the Super gate tuning up a zero bias peak, the authors cropped and zoomed in on one region of their data. Hidden are the criss-crossing states and blobs that are clear in the fuller data and demonstrate quantum dots and disorder, by cropping off significant portions of the scan. They also adjusted the color scheme from the default color scheme for the plotting program, Qtplot (https://github.com/Rubenknex/qtplot). The combined effect was to conceal features that would communicate the presence of disorder and trivial quantum dot states.

As shown below, when we replot the authors' dataset used for Figure 2d without undertaking their undisclosed processing steps, evidence of disordered states, multiple regions of near zero bias states, splitting states, anticrossing and transient states at higher bias voltages, is clear. Please keep in mind that the super-gate is underneath a metal and is heavily screened, so the relatively large gate voltage range shown is not evidence of stable zero bias peaks. Instead, the zero bias peak range in gate voltage should be compared to the scale of the features surrounding the peak. If the two are of the same scale that suggests they are of the same origin. This requires being able to see the features which we present below.

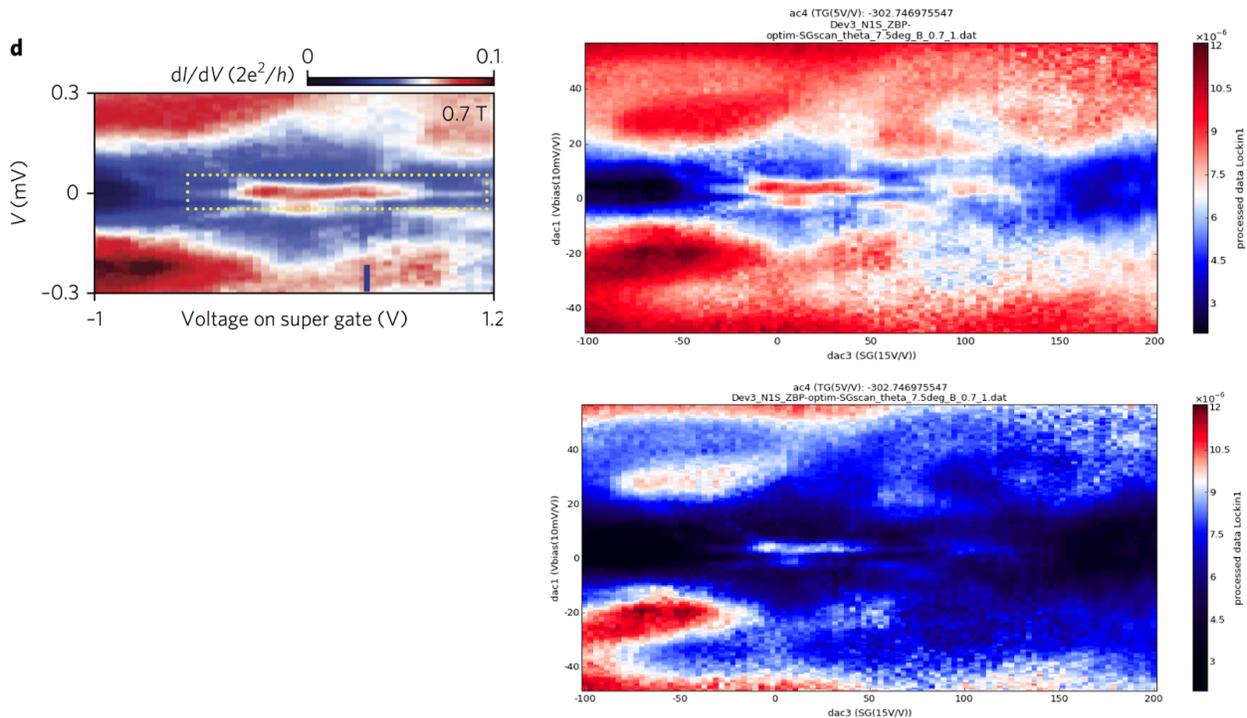

*Replotting of Figure 2d from [1] without cropping, compared with two different versions of the color palette (the first being the standard for the plotting program). It is standard practice in Delft to vary the palette with a single slider control in order to reveal features at different signal magnitudes. Readers typically trust that a palette differing to the default was chosen to reveal additional features; and certainly not to conceal important features undermining the stated conclusions.*

Even prior to these undisclosed data adjustments, the authors had made severe selections to a scan within their full data that was most amenable to this treatment. In other scans in the fuller data, we see that only about one fifth of the gate voltage range is occupied by a zero bias peak, which however comes in several disconnected regions, typical for Andreev bound states. Even blurred diamond-shaped blue regions can be observed, typical of quantum dots with low barriers.

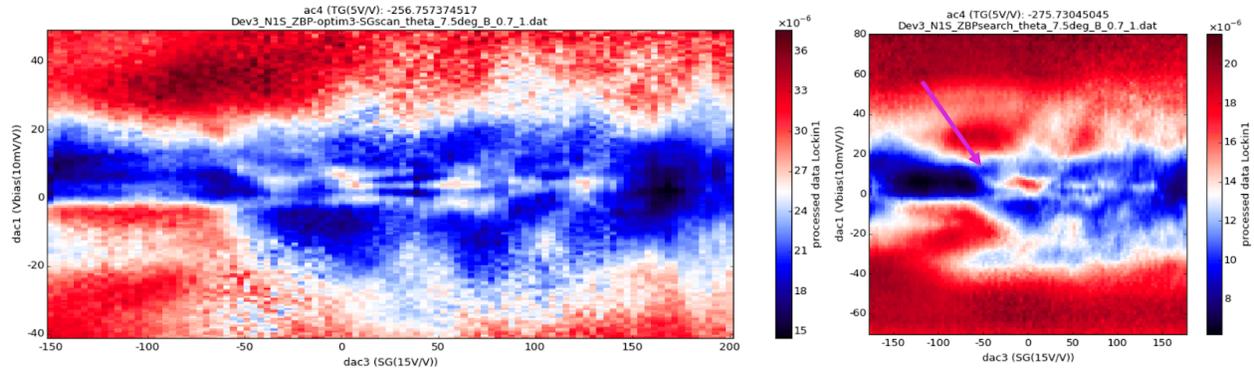

*More super-gate data at B=0.7 from "Device 1". In the illustration above, the scan on the right exhibits a trivial subgap "Andreev" state originating from a quantum dot, shaped as a parabola with a big bright blob at the bottom, indicated by the purple arrow.*

Looking at more scans from this series, we observe that the higher field behavior is revealing more and more clearly the quantum dot features and disorder. These data were not included in the Nature Nanotechnology paper. Focusing on B = 0.8T to B = 1.1T, within the "yellow box" that boasts a simple-looking zero bias peak in Figure 2a, the super-gate scans become more and more disordered with the higher field, as expected when more and more disorder-induced trivial states disperse around in the parameter ranges.

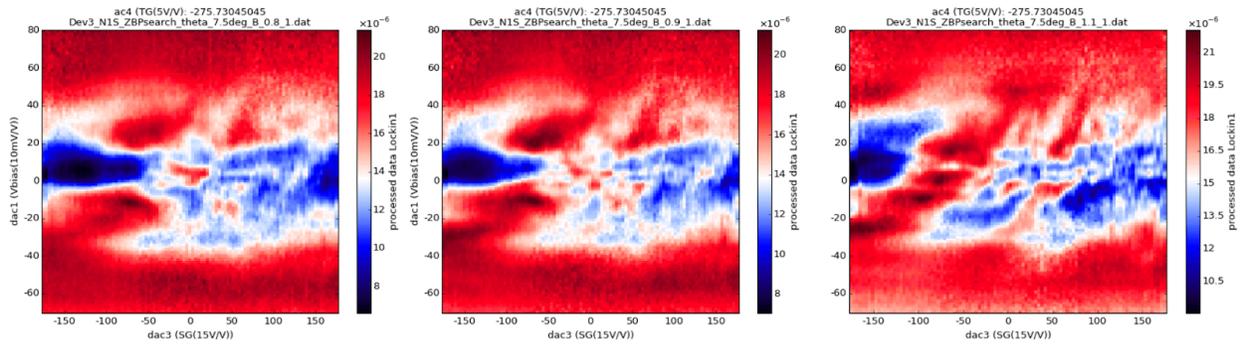

*Super-gate series for fields B=0.8 through B=1.1, within the "yellow box" range where a long and "clean" ZBP can be seen in Figure 2a of the main text. To create an image like Figure 2a one would need to pick a very particular setting of the super-gate where that zero-bias peak extends to large magnetic fields.*

To create an image like Figure 2a one would need to pick a particular setting of the super-gate where that zero-bias peak extends to large magnetic fields. Indeed below are examples at Super Gate of +480 mV, +690 mV, within the bounds of the yellow box in panel 2d, and not far from the setting in panel 2a which is +750 mV, but where the zero bias peak is rather short-lived. Other field sweeps provide further examples of complex near zero bias behavior.

In summary, Figure 2 gives the impression that the zero bias peak is extensive in super-gate voltage, with the idea that this holds for any field up to 1.2T. But the fuller data go against the claims of ballistic superconductivity and of devices any cleaner than others from the same project in earlier time periods, in which near-zero bias features were repeatedly ascribed to trivial non-Majorana states.

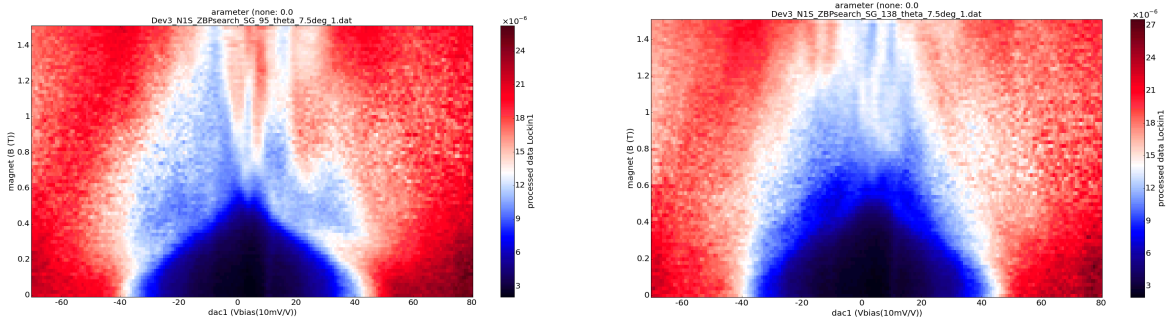

*Magnetic field sweeps at Super Gate of +480 mV, +690 mV*

Figure 3

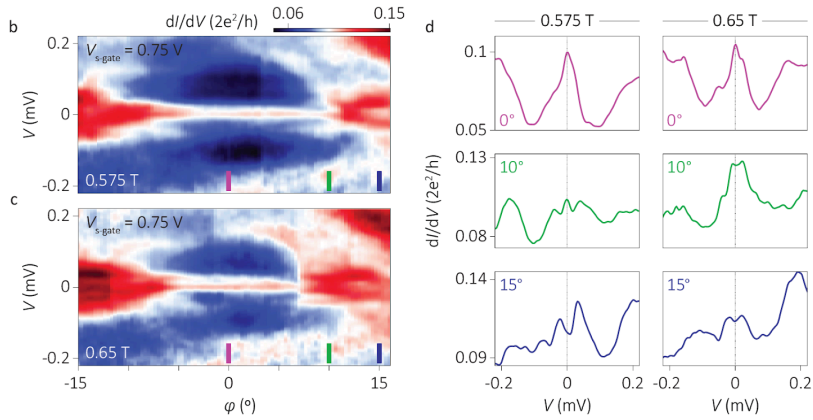

Supplementary Figure 2 | **Dependence of the zero bias peak on the orientation of an in-plane magnetic field. a,** Differential conductance $dI/dV$ as a function of bias voltage $V$, and in-plane rotation of the magnetic field with a magnitude of 0.6 T. $\varphi = 0°$ corresponds to an external field along the wire, $\varphi = \pm 90°$ to an external field parallel to the spin–orbit field $B_{SO}$. The zero bias peak is present in an angle range ($|\varphi| < 10°$) when the external magnetic field is mostly aligned with the wire. We observe a low conductance region around the zero bias peak, indicating the induced gap. Orienting the magnetic field away from the wire axis and more towards $B_{SO}$ closes the induced gap and splits the zero bias peak. We do not observe a stable zero bias peak for $|\varphi| > 10°$ in the entire angle range. The dashed square indicates the angle range shown in main text Figure 3c. **b, c,** $dI/dV$ as a function of $V$, and in-plane rotation of the magnetic field with two different magnitudes. Increasing the magnetic field decreases the angle range of the zero bias peak (compare **b** and **c**). **d,** Vertical line cuts from **b** and **c** at the angles indicated with colored bars. Top panels: For $\varphi = 0°$ the zero bias peak is present for both magnetic field magnitudes. Bottom panels: For $\varphi = 15°$ no zero bias peak is present for both magnitudes. Middle panels: For $\varphi = 10°$ the zero bias peak is present only for 0.575 T, while is split for 0.65 T.

*Figure S2 reproduced from Gül et al., Nature Nanotechnology 2018 [1]*

The authors present Figure 3 as further strong evidence for the Majorana interpretation. Theory predicts that the zero bias peak should vanish when the magnetic field is turned away from the direction of the nanowire, if we are dealing with a zero bias peak of Majorana origin. The paper makes the claim explicit that for angles beyond 10 degrees there is no "stable zero bias peak." We highlight it in the caption of Supplementary Figure 2 that contains more data of the same kind. Two magnetic fields settings are shown in Figures S2b and S2c where from the figures this appears to be true. Zero-bias peaks only found in the narrow window around zero angle are expected for Majorana modes.

This crisp picture is a result of data selection. In the data we found in the 2023 data release, the zero bias peak persists beyond 24 degrees, more than double the maximum angle that the paper claims. Similar larger angle range zero bias peaks can be found at other fields, e.g. B=0.8T and B=0.9T, all below B=1.2T the range of the yellow box in Figure 2. The zero bias peak range is also asymmetric with respect to phi = 0, and this is also inconsistent with the simple Majorana picture in the paper. Finally, the uncropped plots reveal significant disorder, blobs and criss-crossing states indicative of unintended quantum dots.

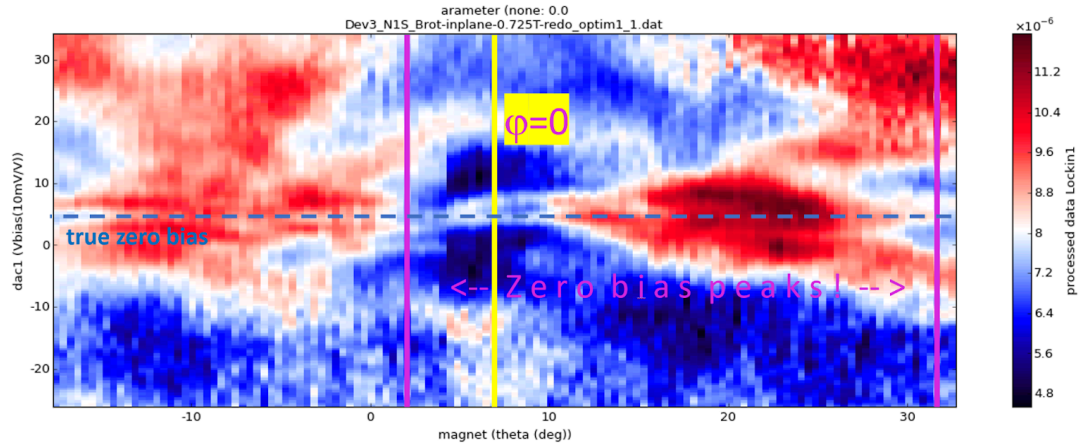

*Sweep of in-plane field angle, at B = 0.725T close to the field amplitude shown in Figure 3. Vertical magenta lines mark the range of ZBP and the zero-angle line. According to the as plotted labels, the ZBP is seen till 31.7 degrees, but the relevant scans for this device used for Figure 2 appear to be taken at an angle of 7.5 degrees which is probably the actual angle of the nanowire with respect to the magnet, and the authors likely refer to that point as phi = 0 in the paper.*

Like Figure 2, Figure 3 also resulted from non-standard color adjustment and cropping. We find evidence in the full data that the authors cropped angle-dependent data to produce Figure 3c and adjusted the color scale to conceal features incompatible with the claims of the paper, or complicating the discussion. The authors cropped the bias voltage range (vertical axis) to minimize the appearance of big blobs at finite bias. They also removed a region of zero bias peak on one end of the data (see pink box in the panel below). And from these data we do not know if that zero bias peak continues to even larger angles. This region poses problems - because the expectation for Majorana-related zero bias peaks is that they live in a narrow vicinity of zero-angle-fields aligned with the nanowire, and then disappear. A re-appearing peak is not consistent with the straightforward Majorana claims of this paper.

Armed with additional data that we obtained in 2023, we find overwhelming evidence that Main Device 1 is a highly disordered nanowire, which should have undoubtedly lead to the conclusion that any near-zero bias features in this device originate from accidental quantum dots induced by disorder and not from Majorana bound states.

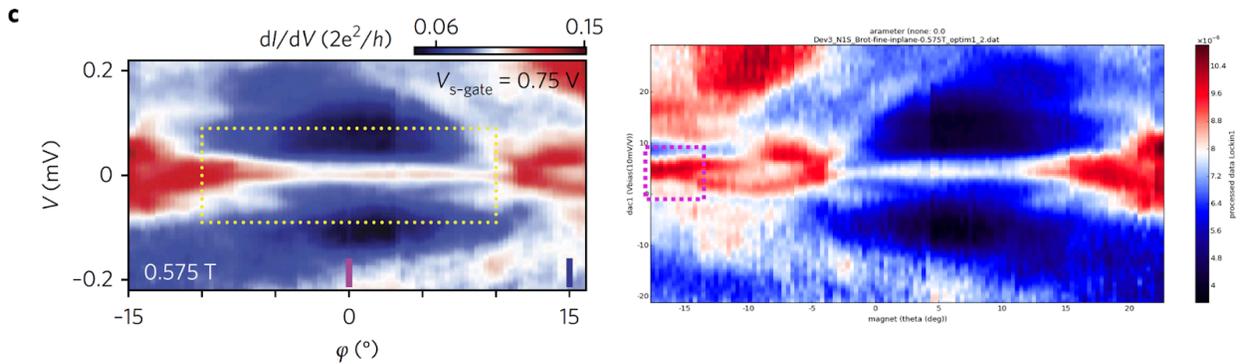

Fig S2b: Same data as presented in the paper (left, from [1]) and uncropped (right). The figure in the paper has the horizontal axis multiplied by -1, the reason for this is unknown but is not a concern.

Figure 4: analysis of "another" device

Towards the end of the paper the authors switch to showing a second device, which they term "another" device, and describe as being "identical" to the device that occupies the narrative of Figures 1-3. Although this device is similar in design to the main device, it was actually made and studied a year earlier than the main device, and is unlikely to be "identical."

The paper describes this figure in simplistic and categorical terms: *"A disorder-free Majorana theory model with parameters extracted from this device (geometry, induced gap, spin–orbit coupling, temperature) finds perfect agreement between simulation and our data."*

The authors present this device as confirmatory. First, the authors re-iterate some of their claims when discussing this device. They claim that it is ballistic, even mentioning the 1.2 micron section covered by the superconductor, to suggest it is ballistic throughout that section. Second, they show new aspects, such as the zero bias peak phase diagram that we analyze below. Finally, they show some of the most visually striking data of the entire paper.

For other devices the authors feature in their "ballistic" papers, their main argument for why these devices are superior are the so-called "pinch-off" data, which are two-dimensional maps of conductance in the space of bias voltage and gate, taken at zero magnetic field. Unfortunately the authors still have not shared analogous tunnel gate scans from this device used in Figure 4. We did plot tunnel gate scans from other devices on the same chip as "another device", which are available. Those data look clearly disordered, and disprove the claim of ballistic transmission over the tunnel gate. Of course, even if the tunnel gate region was ballistic, this says very little about the 1.2 micron long section underneath the superconductor. For that one needs to study the super-gate scans with high resolution, but such data are also missing for "another" device.

The data the authors present in Figure 4 originates from a sequence of 11 fairly large scans (over 10MB each) taken over the course of 3 days in October 2015. Two of these are in panels 4a and 4d. These happen to be the only two datasets from the series which can be used to convey the impression of a simple yet gate-tunable zero bias peak. All other sets from the series of 11 have additional regions of zero-bias peak, charge jumps or show no change with gate compared with the two chosen panels. Even for the two panels, color palette adjustment obscures complicated features.

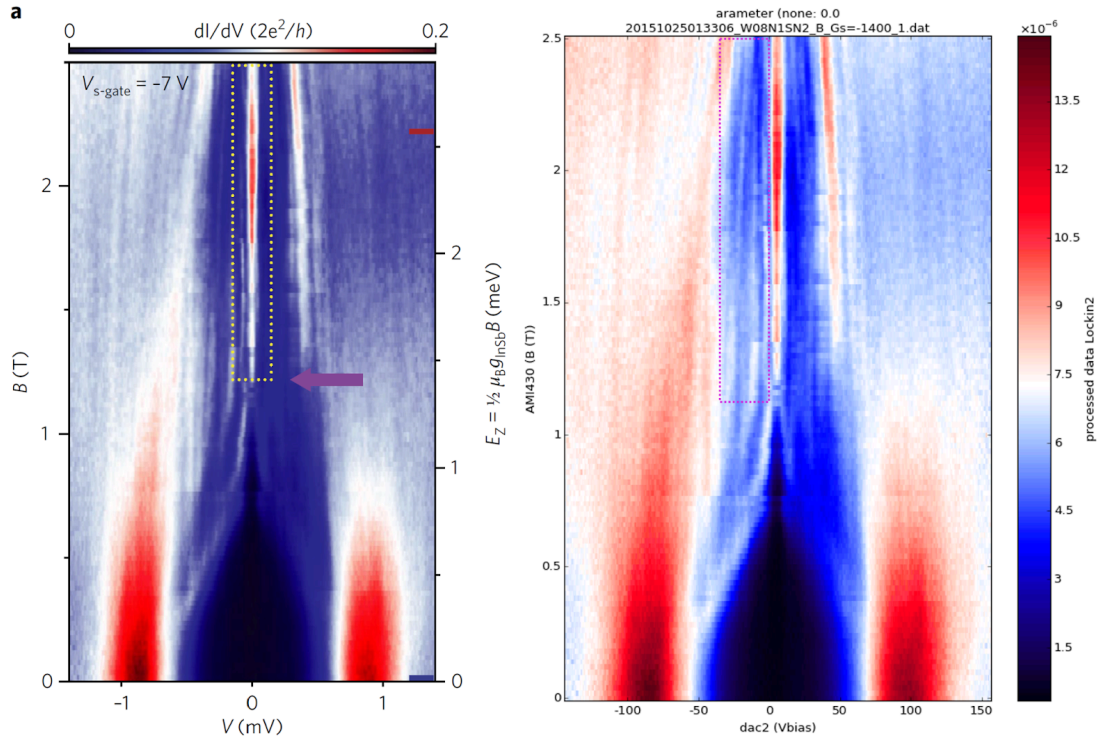

Fig 4a: on the left, as shown in Gül et al Nature Nanotechnology 2018[1], on the right, replotted with a more balanced color scheme, low energy states become more visible in the purple box. Note this is the very same dataset featured by Kouwenhoven in his June 2024 perspective [6].

Topological gap illusion. As presented in the paper, Figure 4a is Majorana-like: long-lived and robust zero-bias peaks that are isolated from other states. One thing we notice when we adjust the colors slightly is that the central (zero-bias) peak is in fact surrounded by numerous peaks inside the deep blue region, "the gap". It is in fact not an isolated peak. Majorana-like appearance suffers when plotted this way, because a zero-bias peak inside a nearly featureless gap strongly hints at a special origin - nothing keeps that peak at zero bias except, perhaps, topology of the bands. While if we see many states next to that peak, those states can be pushing on the peak, due to level repulsion, the effect discussed in the literature as early as 2014. The question of topological gap is central to any quantum computing prospects of this research.

The authors also present, in panel 4e, a "phase diagram" of zero bias peaks. This is another way to assess how likely the Majorana interpretation is, as well as to gain assurance that the authors have performed comprehensive checks of their data. For each of the 11 large magnetic field scans, they write that they extracted the fields for the onset and the disappearance of the zero-bias peaks. They then connected the onsets and the disappearances by lines and grayed in the region in between, labeling it "zero bias peaks." No further description of their process is included in the main text or supplementary, though all 11 data sets are plotted in the supplementary in small panels.

Having obtained the original 11 datasets, we replotted them. We find that the phase diagram presented in Figure 4e omits multiple regions of zero bias peaks present in the data, which the authors have excluded and not labeled in Figure 4e. The approximate positions of those intervals are labeled in our figure with orange solid bars. There is no justification for leaving out these zero bias peak regions.

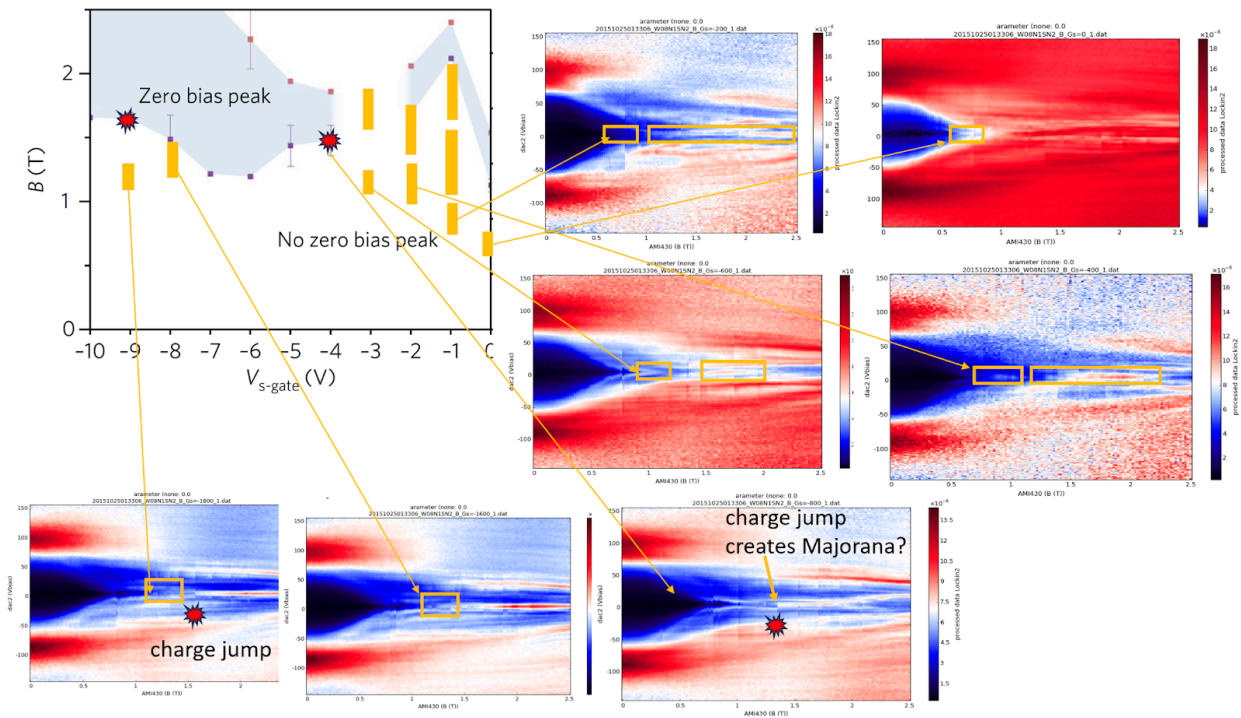

*In this Figure, we show with orange bars the zero-bias peaks not labeled in the phase diagram Figure 4e of Gül et al Nature Nanotechnology 2018[1]. Red stars mark the positions of charge jumps. The orange rectangles mark where we found zero-bias peaks when we replotted the data.*

With the omitted zero bias peaks plotted, this important panel fails to support the authors' Majorana claims, at least in the form made where the simple Majorana theory fully explains the data. The complete plot reveals zero bias peak regions that are disconnected, which suggests they come from other states coming down to zero bias, which would be disorder-induced. But the paper claims there are no such states in any of the devices. Some of the peaks excluded are fairly tall while others are more faint, yet no weaker than those reported in Figure S6, and certainly above the noise.

In their 2021 data release the authors provide a Readme file in the S3 folder in which they reveal their knowledge of these extra zero bias peak regions and comment on their removal. However, no such information was provided with their published paper. The authors claim they removed these zero bias peaks because they were "not robust," which is what anyone would expect from trivial disorder-induced states; the presence of which the paper denies.

Another important observation is that in two of the 11 datasets, zero bias peaks appear/disappear at the location of a charge jump. We mark these instances with red stars above. The implications of this for the Majorana interpretation are dire. A single charge instability controls the topological nature of a 1.2 micron long segment of the nanowire. This further enhances the likelihood that this zero bias peak derives from a local quantum dot.

## Analysis of Additional Device 1

Additional Device 1 features in the supplementary information Figures S5 and S6. The main text states that key observations of ballisticity and zero bias peak robustness from "Main Device 1" are replicated in this device. In this capacity, the device plays a very important role in the paper. Both claims are not true and the confirmatory appearance of these Figures relies on undisclosed data processing.

First, we already identified in 2021 that this device does not demonstrate ballistic behavior. In Figure S5, features were cropped beyond the one plateau-like segment, because they looked like more of the same kind of plateaus, which however couldn't be aligned with the expected quantized values regardless of the chosen arbitrary series resistance offset.

Second, the authors omitted significant evidence that what they presented as a robust zero bias peak in Figure S6 is a clearly resolved trivial level crossing without any persistent zero bias peaks. Both types of data omission and undisclosed processing render the research inaccurately presented.

## Unreliable ballistic conductance claim

Figure S5 is supposed to present confirmation of the evidence in Figure 1; it shows similar data from another device. Likewise there we see a black trace in panel S5d which is attributed to a quantized plateau. In this Figure, the undisclosed contact resistance of 500 Ohms was subtracted, as opposed to 3 kOhms for Figure 1. Furthermore, in the cropped out region of the data, at gate voltages more positive than 1.3V, there is another flat region of conductance which appears similarly plateau-like as the one shown in Figure S5d, except the value of conductance does not correspond to any quantized value. Moreover, if a gate voltage trace is selected at a different bias voltage, both features evolve and shift around in conductance to different values, neither of which appears pinned to a quantized conductance. So if another bias voltage were to be selected for Figure S5d, the Figure would look different, and require a subtraction of a different offset to position some feature at the value of 1.

Figure S5 also contains a cut-out segment in the middle of the data, an adjustment not disclosed in the paper. The meaning of "cut-out" is that a segment of data - in this case 12 vertical traces of bias voltage - were removed from the dataset. Upon their removal, the data to the right, with more positive gate voltages, were shifted so that different gate voltages were assigned to each current-voltage trace. These new gate voltages were not the ones at which the data were obtained.

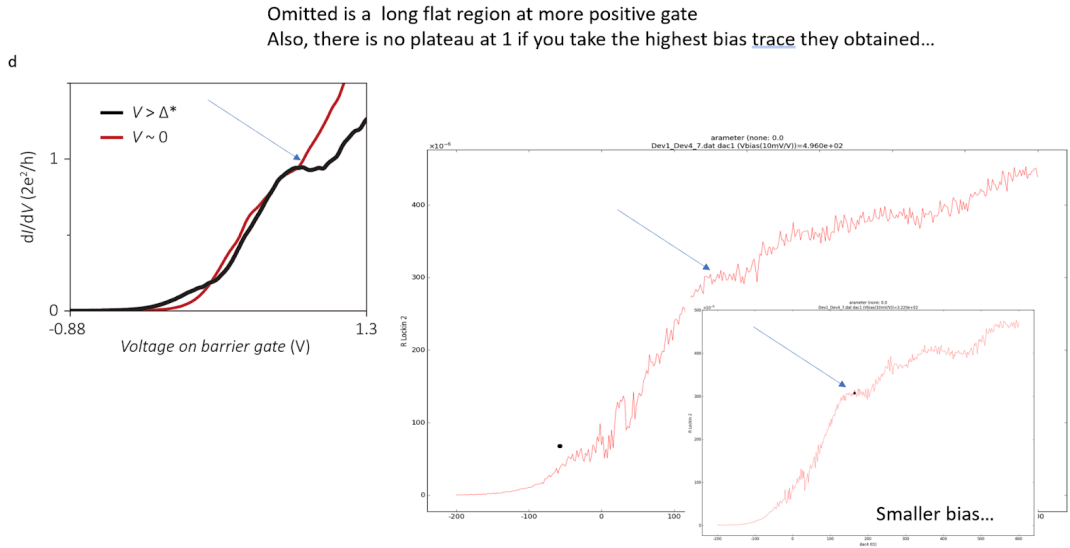

The same problem of the authors showing too narrow a voltage bias range undermines their claims about this device. In the colorscale we can also see that a range of gate voltage and bias voltage was missing from the paper. If the full dataset were shown it would have created questions as to whether any evidence of quantized conductance exists for this nanowire.

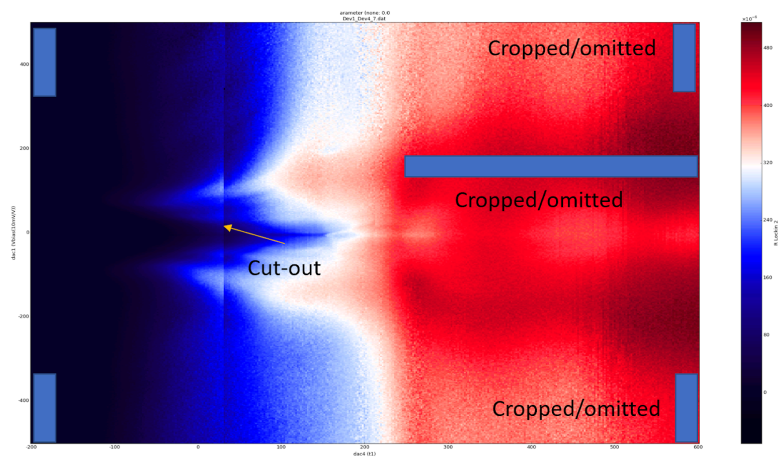

Our analysis established that data from Additional device 1 did not contain evidence for quantized conductance. While the more technical manipulations were the subject of a recent correction, the cropping-out of the flat regions at larger gate voltages that cannot be aligned to any quantized values remains an unaddressed and a glaring problem. Such flat regions are readily explained by the presence of weakly confined unintentional quantum dot(s) that evolve into quantum interference patterns at higher conductance.

No robust zero-bias peak in Additional Device 1

This part of the analysis focuses on Figure S6. The caption title states "*Additional device 1 - zero bias peak in a large range of magnetic field and local gate voltages*," conveying that the zero bias peak is very robust and Majorana-like.  In discussing panel S6a (left panel below), the authors claim the zero-bias peak extended in magnetic field over 400 mT: "*Application of a magnetic field generates a zero bias peak at 0.3 T. The zero bias peak persists up to 0.7 T and splits for larger magnetic fields*,[..]". This is reinforced through labeling the right-hand axis in terms of Zeeman energies (E_Z) with an effective g-factor of 40, not unreasonable for InSb.

Upon studying additional data made available in 2023, we discover that there is no zero bias peak in these data, but a crossing of two states that do not stick to zero bias at all. To show this, we first replot the dataset shown in S6a. We choose an appropriate color scale that does not hide features, and then we can see a pair of peaks around zero bias near 0 mT (middle panel below).

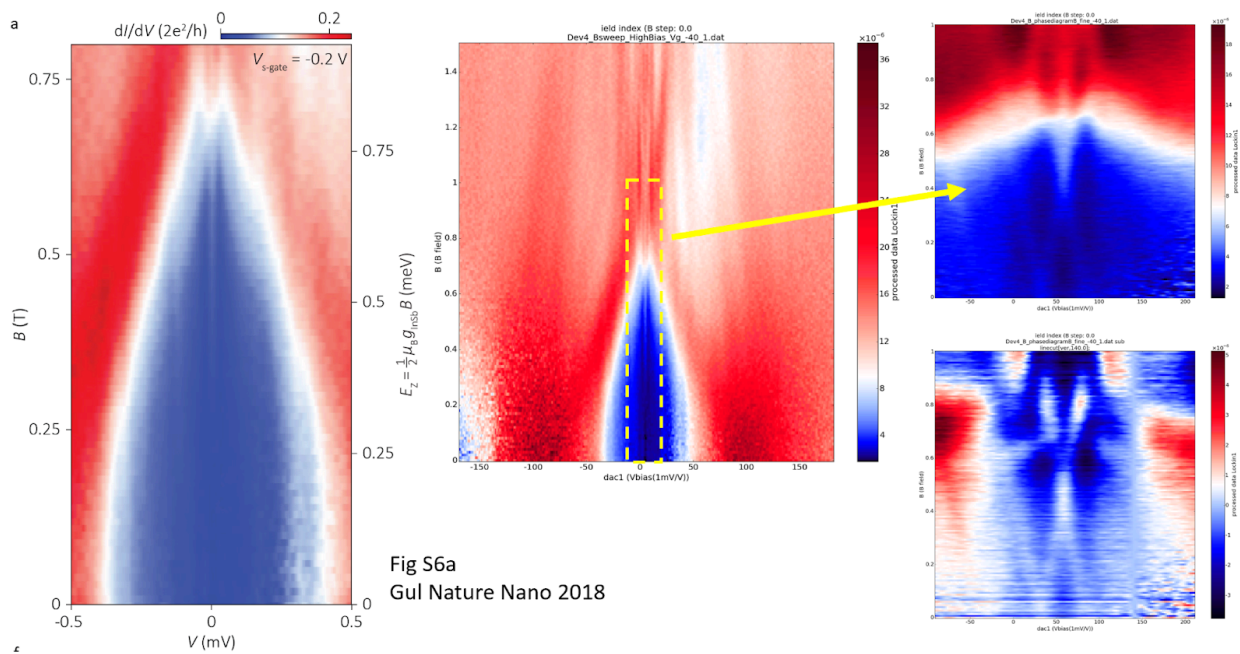

Although the authors did not say so in their paper, they investigated this particular pair of states in great detail. They recorded over 150 datasets. In many of these labeled "fine" they reduced the bias step size and narrowed the range of bias voltages to +- 150 microV to look carefully at features near zero bias. We located their high resolution dataset at the Super-gate used to generate Figure S6a (yellow arrow above). At zero magnetic field, we see finite bias states at approximately +- 25 microV. In a magnetic field, these states spin-split and linearly disperse. This was by 2018 already a textbook style example of a spinful Andreev bound state. There is no visible deviation from linear dispersion, or any zero-bias sticking behavior. To simplify the g-factor extraction, we subtracted a background signal defined by the vertical linetrace at +140 microV from the entire dataset (bottom right panel).

We estimate the effective g = 2.3. Scan after scan in this large series shows the same linearly dispersing states and no robust zero-bias peaks. **With such a low g-factor, not only is this not relevant to Majorana, it is not relevant to InSb that the nanowire is made of.** Low g-factor states dispersing in a magnetic field were already discussed in the Mourik et al., Science 2012 paper from the same group[9] (see supplementary materials there).

Despite having acquired the high resolution scan at the same gate voltage, and many other scans in this regime, the authors picked the low resolution dataset for S6a. The choice obfuscates the low energy features. The authors also stretched the low resolution scan vertically to further visually de-emphasize any dispersion of their low energy features. They also manipulated the colorscale so as to make the initial spin splitting non-detectable. They included a large effective g-factor axis in their panel. This was all contrary to the fact that their claimed robust zero bias peak is a level crossing of a pair of states dispersing with an effective g-factor of about 2.

We conclude the authors made undisclosed alterations in this figure, to support their claims of ballistic behavior and robust zero peaks compatible with Majorana modes. In their own high resolution data, the low bias peak reveals the presence of an impurity state, and their broad resonances present in the pinch off show that weakly confined quantum dots present in the device.

Analysis of Additional Device 2

The same problem of data shared only in 2023 revealing evidence of disorder, especially at finite magnetic fields, repeats itself when we plot more data from "Additional Device 2" which was included in the supplementary information Figure S7.

Our plots reveal complicated and disordered regimes, with zero bias features embedded within criss-crossing states and blobs moving around. The figures in the paper itself appear much cleaner. To understand how the authors achieved this, let us focus on the super-gate dependence of the zero bias peak in this device, Figure S7e. When we plot the data we see a short-lived zero bias peak surrounded by charge jumps and disorder features; the peak appears to not persist longer than the criss-crossing resonances. But the version in the paper looks like a long zero bias peak on a quiet background, isolated and extended. We can reproduce this appearance by beginning with the original raw data, then 1) manipulating the aspect ratio by stretching the panel and 2) manipulating the color palette to conceal some of the diversity of features and 3) cropping to a narrow parameter range in bias voltage (y axis) and gate (x axis), to conceal the big red state moving through.

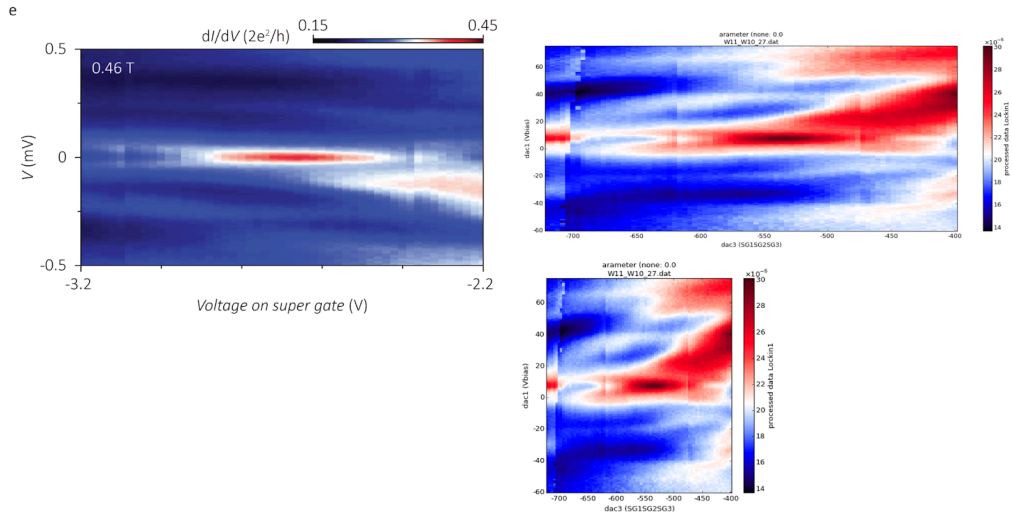

*Stretched and compressed versions of data in Fig S7e from [1]. (The figure in the paper has the vertical axis multiplied by -1, the reason for this is unknown. We do not see a clear issue with this, unless this was done to conceal further the features that interfere with the Majorana-like appearance.)*

Data from which Fig. S7e was derived are part of a series of gate scans at different magnetic fields. The sequence, some of which we plot below, reveals unmistakable disorder features which coexist and interact with the zero bias peak. Presented this way, using the default color scale setting of the plotting script Qtplot (https://github.com/Rubenknex/qtplot), it is hard to argue that this is a disorder and quantum dot free situation. And if this is a disordered regime, then any direct interpretation of these features as related to Majorana is not possible. Note the significance of mentioning the default colorscale is that this is what the person plotting would likely see first on their screen. The authors changed to a less informative colorscale that obscured scientifically meaningful features, cropped and stretched their panels.

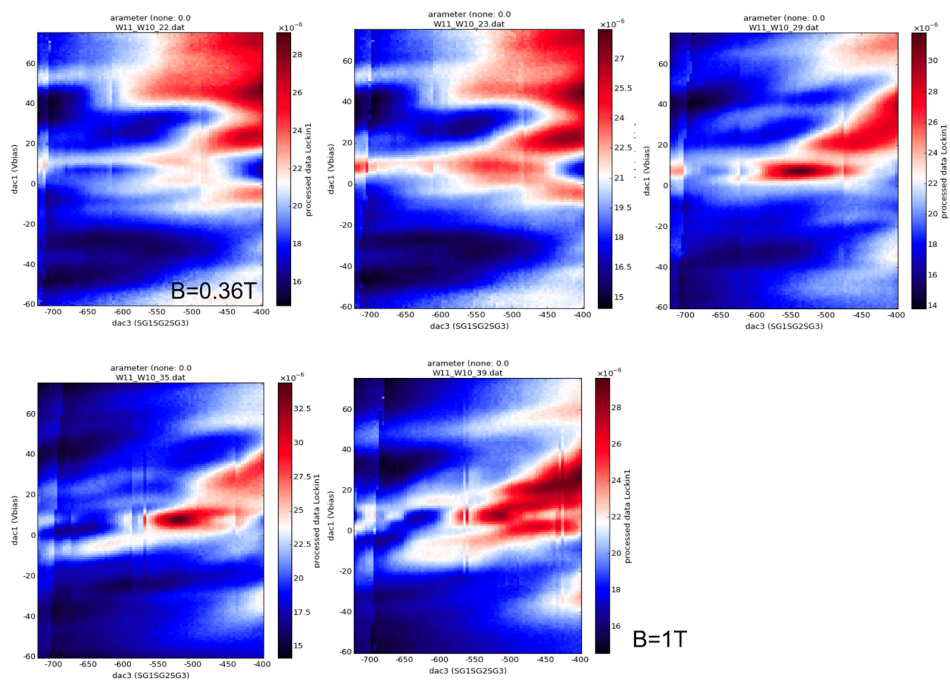

# Further Discussion

## Note on theory

In Gül et al., Nature Nanotechnology 2018[1], and in some of the communications from the authors, they put forward as a discussion point that their theoretical model mimics some of the features of their data. In fact, they praise the agreement. The theory in question is contained in the parallel publication, Zhang et al., Nature Communications 2017[2]. However, it gives unphysical outputs. Such as mean free paths of microns, sometimes exceeding the nanowire length, even though the distance between the contacts is only 70-100 nm. It is unphysical to deduce a mean free path that exceeds by 1-2 orders of magnitude the length of the region over which the majority of current passes. The theory then translates this into ultrahigh transparency of the contact barriers. In principle, such unrealistic outputs should trigger a deeper inspection, a discussion in the paper and a re-evaluation of the model used. Or, at the minimum, any theory discussion should disclaim the numerical accuracy of model parameters. But here the authors simply embrace these huge numbers as evidence of how excellent their devices are.

The biggest problem with the model is that it does not include quantum dot effects which can present themselves as plateau-like features and lead to enhancements of current when a dot is coupled to a superconducting lead. Quantum dots also host trivial Andreev states that exhibit robust zero-bias peaks. Theory is used as a circular argument, not including quantum dots in the model and then finding that they are not present.

## Outside gates

Figures 2b and 2c show dependences on outside gates. The authors cropped these data to avoid presenting charge jumps, a pattern present throughout a four-paper series from Delft. However, the quantum dot or dots they were looking at is clearly sensitive to the Super-Gate, and thus it is located there, underneath the superconductor. It is not hard to understand how the authors were able to generate some plots for the "Barrier gate" and the "Normal gate" that appear to contain fewer features of disorder, such as seen above in their Figure 2, in Figure 1 or indeed in their Nature Communications 2017 paper[2]. The authors chose parameter ranges that are very narrow compared to the scale of the relevant quantum dot features for those gates specifically. Some of this is achieved through oversampling and aspect ratio adjustment (i.e. stretching of panels), while another factor at play for those less relevant gates is the chosen geometry of their devices.

The contact metals N and S are fully covering the nanowire (above and below the wire in the SEM image in Figure 1), and the distance between N and S, the space where the barrier gate is located, is very narrow, the same as the wire diameter. In this configuration, screening of the electric fields from the gates results in the expanded range of gate voltage needed to change the signal. This makes it look like the gates were changed by a lot, in terms of Volts applied.

At the same time, the electric field profile is very sharp; the opening of the barriers in this configuration happens more rapidly than the tuning of the multiple quantum dots under the superconductor, creating the illusion of a more monotonic trace. These phenomena are well known to the authors: indeed the previous generations of devices from the same team going back to 2012 had electrodes spaced farther apart and half-covering the wire to enhance the gate effect, so the Delft group has experience working with very different gate designs. Still, for assessing whether this wire has any promise for a qualitatively better observation of Majorana (whether it is ballistic) it is important to assure that there are no quantum dots, rather than find ways to obtain and process data where quantum dot signatures are less apparent.

One tool that the experimentalists on this project had to assess the disorder in their samples in such an extreme geometry, where the nanowire is surrounded by metals, was to switch on the magnetic field. They apparently knew this as they indeed acquired such data but did not share most of it in the paper, or as part of their 2021 data release. The importance of the magnetic field tuning is its capacity to unpin trivial subgap "Andreev" states which tend to be less visible and bunch together at zero field, and spread them randomly through the space in gate voltage and bias voltage, making them more visible. Even if you don't see such features at zero field, but then you do see them at finite fields - the conclusion is that quantum dots are actually present in the device, and thus a clean Majorana interpretation is not possible.

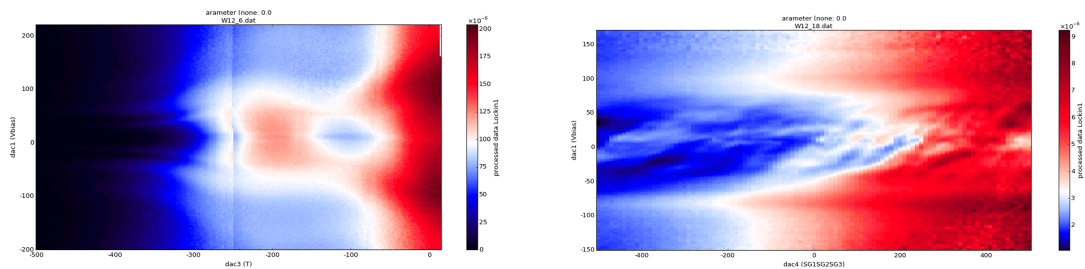

*Left: zero-field tunnel gate data on a device not included in the Figures of the paper. Right: finite magnetic field super-gate data.*

Here is an example of this behavior, from a nanowire that the authors did not include in the Nature Nanotechnology paper, but that was measured at the same time. Zero field data present itself as more smooth, though the red blobs are quantum dot resonances and not ballistic "Andreev reflection enhancement" effects that the authors claim in their papers. These are oversampled plots taken over a narrowed window of gate and bias voltages.

But the finite field "super-gate" data look very disordered. This is at 0.5T, and we see numerous low bias states, some forming quasi-persistent zero-bias peaks and some dispersing through. If you see this type of behavior in several devices, you cannot simply pick data where this is less apparent as Figures for your paper. We also provide these data to illustrate that there is more data, from more devices, and our analysis of those data is that they consistently reveal significant disorder.

## Conclusion

The 2023 data release has exposed evidence of undisclosed data adjustment and processing that obscures the fact that the zero-bias peaks shown in the paper are of trivial quantum dot origin. This evidence exists in previously entirely unpublished data, and in published data but in regions of parameter space only just beyond the very narrow ranges to which data were cropped for the paper, and in the ranges shown, were it not for an idiosyncratic choice of color scale. In our view, such data presentation choices undermine the basis for the paper.

In our view, the authors' undisclosed observations of quantum dots and non quantized conductance alone invalidate their claim of "ballistic" devices, thereby severing the link between the zero-bias peaks they saw and the possibility of Majorana physics. The fact that zero-bias peaks in the 2023 data exhibit many instances of non-Majorana behavior typical for quantum dot-borne trivial Andreev states definitively voids the authors' claims that any of their data are due to Majorana.